\documentclass[twocolumn]{aastex631}

\newcommand{\G}{\mathcal{G}}

\newcommand{\appropto}{\mathrel{\vcenter{\offinterlineskip\halign{\hfil$##$\cr\propto\cr\noalign{\kern2pt}\sim\cr\noalign{\kern-2pt}}}}}

\newcommand{\teff}{\ensuremath{T_{\rm eff}}}
\newcommand{\logg}{\ensuremath{\,{\rm log}\,{g}}}
\newcommand{\ecosw}{\ensuremath{e\cos{\omega_*}}}
\newcommand{\esinw}{\ensuremath{e\sin{\omega_*}}}
\newcommand{\msun}{\ensuremath{\,{\rm M_\Sun}}}

\newcommand{\mstar}{\ensuremath{\,M_{\rm *}}}
\newcommand{\rstar}{\ensuremath{\,R_{\rm *}}}
\newcommand{\mj}{\ensuremath{\,{\rm M_{\rm J}}}}
\newcommand{\prot}{\ensuremath{\,{P_{\rm rot}}}}

\newcommand{\degree}{\ensuremath{\,^{\circ}}}
\newcommand{\mplanet}{\ensuremath{\,M_{\rm P}}}
\newcommand{\rp}{\ensuremath{\,R_{\rm P}}}

\newcommand{\vsini}{\ensuremath{v\sin{i_*}}}

\usepackage{url}

\usepackage{amsmath}
\usepackage{float}
\usepackage{rotating}
\usepackage{hyperref}
\graphicspath{{./}{Figures/}}

\shorttitle{The Spin-Orbit Misalignment of TOI-1842b}
\shortauthors{Hixenbaugh et al.}

\begin{document}
\title{The Spin-Orbit Misalignment of TOI-1842b \\
\small{The First Measurement of the Rossiter-McLaughlin Effect for a Warm Sub-Saturn around a Massive Star}}

\author[0000-0002-8685-5397]{Kyle Hixenbaugh}
\affiliation{Department of Astronomy, Indiana University, Bloomington, IN 47405, USA}

\author[0000-0002-0376-6365]{Xian-Yu Wang}
\affiliation{Department of Astronomy, Indiana University, Bloomington, IN 47405, USA}

\author[0000-0002-7670-670X]{Malena Rice}
\altaffiliation{51 Pegasi b Fellow}
\affiliation{Department of Physics and Kavli Institute for Astrophysics and Space Research, Massachusetts Institute of Technology, Cambridge, MA 02139, USA}
\affiliation{Department of Astronomy, Yale University, New Haven, CT 06511, USA}

\author[0000-0002-7846-6981]{Songhu Wang}
\affiliation{Department of Astronomy, Indiana University, Bloomington, IN 47405, USA}

\correspondingauthor{Kyle Hixenbaugh}
\email{khixenb@iu.edu}

\begin{abstract}
The mechanisms responsible for generating spin-orbit misalignments in exoplanetary systems are still not fully understood. It is unclear whether these misalignments are related to the migration of hot Jupiters or are a consequence of general star and planet formation processes. One promising method to address this question is to constrain the distribution of spin-orbit angle measurements for a broader range of planets beyond hot Jupiters. In this work, we present the sky-projected obliquity ($\lambda=-68.1_{-14.7}^{+21.2} \, \degree$) for the warm sub-Saturn TOI-1842b, obtained through a measurement of the Rossiter-McLaughlin effect using WIYN/NEID. Using the projected obliquity, the stellar rotation period obtained from the \emph{TESS} light curve, and the projected rotation velocity from spectral analysis, we infer the 3D spin-orbit angle ($\psi$) to be $\psi=73.3^{+16.3}_{-12.9}\degree$. As the first spin-orbit angle determination made for a sub-Saturn-mass planet around a massive ($\mstar=1.45 \, \msun$) star, our result presents an opportunity to examine the orbital geometries for new regimes of planetary systems. When combined with archival measurements, our observations of TOI-1842b support the hypothesis that the previously established prevalence of misaligned systems around hot, massive stars may be driven by planet-planet dynamical interactions. In massive stellar systems, multiple gas giants are more likely to form and can then dynamically interact with each other to excite spin-orbit misalignments.
\end{abstract}


\section{Introduction} 
\label{section:intro}
Observed trends in stellar obliquity, the angle between a star's spin axis and the net orbital angular momentum vector of its companion planets, have provided insights into the prevalence of different formation mechanisms that shape the demographics of exoplanet systems \citep{schlaufman2010evidence, winn2010hot, albrecht2012obliquities, wang2021aligned, albrecht2021perpendicular,rice2022HJobliqDistr}. Measurements of the Rossiter-McLaughlin \citep[R-M,][]{rossiter1924detection, mclaughlin1924some} effect have revealed that a significant fraction of hot Jupiters are spin-orbit misaligned (see \citealt{winn2015} and \citealt{albrecht2022obliq_review} for comprehensive reviews). However, the origins and evolution of spin-orbit misalignments remain unclear. 

Perhaps the most compelling observational pattern in stellar obliquity measurements thus far is the discovery that hot Jupiters around cool, low-mass stars are preferentially aligned. In contrast, hot Jupiters around hot, massive stars span a wide range of spin-orbit angles \citep{winn2010hot, schlaufman2010evidence, albrecht2012obliquities, wang2021aligned}, known as the $\teff$ vs.\ obliquity relationship. This has conventionally been explained as a signature of tidal realignment, which operates with higher efficiency in cool, low-mass stars with hot Jupiters. Cool, low-mass stars, which have thick convective envelopes, may realign with the orbital planes of their close-in Jupiter companions through tidal interactions on a timescale $\tau_{\Psi} \propto  (\mplanet / \mstar)^{-2}(a / \rstar)^6$ \citep[e.g.][]{albrecht2012obliquities,albrecht2022obliq_review} that is often shorter than the system lifetime. In such tidal interactions, the planet realigns the outer convective layer of the star to the planet's orbital axis \citep{winn2010hot,albrecht2022obliq_review}.

A growing sample of constraints for parameters such as stellar obliquity and eccentricity has recently enabled further studies examining the robustness of this trend in different parameter regimes. For example, \citet{rice2022HJobliqDistr} demonstrated that the $\teff$ vs. obliquity trend has so far only clearly held for hot Jupiters on circular ($e=0$) orbits.



It remains unknown whether the observed $\teff$ vs.\ obliquity relationship (which can be extended to a $\mstar$ vs.\ obliquity relationship, as stellar mass and temperature are correlated for main-sequence stars) also extends to other planet demographics beyond $e=0$ hot Jupiters. Previous obliquity measurements have focused on hot,
massive planets around both hot and cool stars. Recent progress has been made for measuring obliquity for lower-mass planets and warm planets around cool stars \citep{schlaufman2010evidence,sanchis2013kepler63,wang2018kepler9,anisman2020wasp117,dong2022toi-1268,wang2022wasp148,rice2022WJaligned}. Making R-M measurements for lower-mass planets -- particularly those on wide orbits around hot, massive stars, in a poorly populated region of parameter space -- is vital to test whether the $\mstar$/$\teff$ vs.\ obliquity relationship extends to populations other than $e=0$ hot Jupiters. Testing this will shed new light unto the understanding of the origins and evolution of spin-orbit misalignments.

In this work, we present the fifth result from the Stellar Obliquities in Long-period Exoplanet Systems (SOLES) survey \citep{rice2021k2-140SOLES, wang2022wasp148, rice2022WJaligned, rice2023orbital}: a measurement of the sky-projected spin-orbit angle ($\lambda=-68.1_{-14.7}^{+21.2} \, \degree$) of TOI-1842b \citep{wittenmyer2022toi1842}, a warm ($P=9.5740\pm0.0001 \,$ days) sub-Saturn ($\mplanet=0.19_{-0.04}^{+0.06}\mj$) on a slightly eccentric orbit ($e = 0.13 ^{+0.16}_{-0.09}$) around a massive ($\mstar = 1.45^{+0.07}_{-0.14} \, \msun$) star. This is the first Rossiter-McLaughlin measurement of a sub-Saturn-mass ($\mplanet<0.3 \, \mj$) planet around a high-mass star ($\mstar>1.2 \, \msun$). We observed the R-M effect using the NEID spectrograph \citep{schwab2016neid} on the WIYN 3.5 m telescope. The measured obliquity helps demonstrate the framework for an alternative explanation for the origins and evolution of spin-orbit misalignments based on the idea of dynamical excitement \citep{wu2023evidence}. In this framework, initial planet multiplicity is the key tracer of misalignments, and the ability for planet-planet interactions is the key mechanism that is capable of exciting obliquity.

The present paper is structured as follows: the R-M measurement is detailed in Section \ref{section:observations}, followed by a description of the methods utilized to acquire the stellar parameters for the studied system are outlined in Section \ref{section:stellar_parameters}. In Section \ref{section:spinorbitmodel} the global model employed to measure the planet's spin-orbit angle is described.  Finally, the implications of the results are discussed in Section \ref{section:discussion}.

\section{Observations}
\label{section:observations}
TOI-1842 was observed on May 9, 2022 with the high-resolution ($R \sim 110,000$) WIYN/NEID spectrograph. A total of 20 radial-velocity (RV) measurements were taken during the transit of TOI-1842b, with 1200-second exposure times, from 03:35-10:17 UT. The atmospheric conditions during the observations included a seeing of $0.8-1.4\arcsec$ and an airmass range of $\mathrm{z} = 1.1-2.5$. At a wavelength of $5500 \rm \AA$, the NEID spectrograph had a typical signal-to-noise ratio of 66 pixel$^{-1}$. 

The spectra from NEID were processed using the NEID Data Reduction Pipeline\footnote{More information can be found here: https://neid.ipac.caltech.edu/docs/NEID-DRP/}, and the resulting RVs were obtained from the NExScI NEID Archive\footnote{https://neid.ipac.caltech.edu/}. The NEID RV data obtained in this work is provided in Table \ref{tab:rv_data} and displayed in the top panel of Figure~\ref{fig:rm_fit}.

\begin{deluxetable}{ccc}
\tablecaption{NEID radial velocities during a transit of TOI-1842b.\label{tab:rv_data}}
\tabletypesize{\scriptsize}
\tablehead{
\colhead{Time (BJD$_{\rm TDB}$)} & \colhead{RV (m/s)} & \colhead{$\sigma_{\rm RV}$ (m/s)} }
\tablewidth{300pt}
\startdata
2459708.6678852 & 3244.8 & 2.8  \\
2459708.6819921 & 3239.7 & 2.8  \\
2459708.6960637 & 3244.2 & 2.9  \\
2459708.7099717	& 3242.3 & 3.3  \\
2459708.7249360	& 3246.7 & 3.3  \\
2459708.7391807	& 3240.4 & 2.8  \\
2459708.7528341	& 3244.6 & 2.6  \\
2459708.7664207	& 3257.2 & 3.1  \\
2459708.7803821	& 3256.1 & 4.0  \\
2459708.8041891	& 3251.6 & 3.3  \\
2459708.8184957	& 3250.9 & 3.2  \\
2459708.8331761	& 3248.0 & 3.0  \\
2459708.8466759	& 3250.6 & 2.9  \\
2459708.8609819	& 3246.0 & 3.3  \\
2459708.8761644	& 3252.5 & 3.2  \\
2459708.8894657	& 3247.2 & 2.7  \\
2459708.9040107	& 3240.5 & 2.8  \\
2459708.9179683	& 3236.4 & 2.9  \\
2459708.9325356	& 3238.7 & 3.0  \\
2459708.9458221	& 3235.8 & 3.0  \\
\enddata
\end{deluxetable}

\section{Stellar Parameters} 
\label{section:stellar_parameters} 
We determine the best-fit stellar atmospheric parameters ($\teff$, $\logg$, [M/H], and $\vsini$) for each individual spectrum obtained from NEID for TOI-1842 by applying the \texttt{iSpec} code \citep{blanco2014ispec, blanco2019ispec} with MARCS atmosphere models and the GES atomic line list to generate synthetic spectra. We employed the synthetic spectral fitting technique, which minimizes the chi-square value for the difference between synthetic and observed spectra using a nonlinear least-squares (Levenberg-Marquardt) fitting algorithm \citep{Markwardt2009}. The best fit parameters for each individual spectra are combined into a distribution. We take the median of the distribution as the parameter value, and derive uncertainties from the scatter of the distribution for each parameter.

We then used \texttt{EXOFASTv2} \citep{eastman2019exofast} to perform a spectral energy distribution (SED) fit of TOI-1842
by combining the MESA Isochrones \& Stellar Tracks \citep[MIST;][]{Choi2016mist,dotter2016mist} model with broadband photometry from multiple catalogs including Gaia DR2 \citep{gaia2018}, 2MASS \citep{Cutri2003}, and AllWISE \citep{Cutri2013}. We applied Gaussian priors to $\teff$ and [M/H] based on the values obtained from the \texttt{iSpec} analysis, as well as the corrected stellar parallax from Gaia DR2 \citep{Stassun2018}. We also enforced an upper limit on $V$-band extinction using Galactic dust maps \citet{schlafly2011dust_maps}. The resulting stellar parameters are presented in Table \ref{table:results}.

\section{Obliquity Modeling} 
\label{section:spinorbitmodel}

\begin{figure}
    \includegraphics[width=\linewidth]{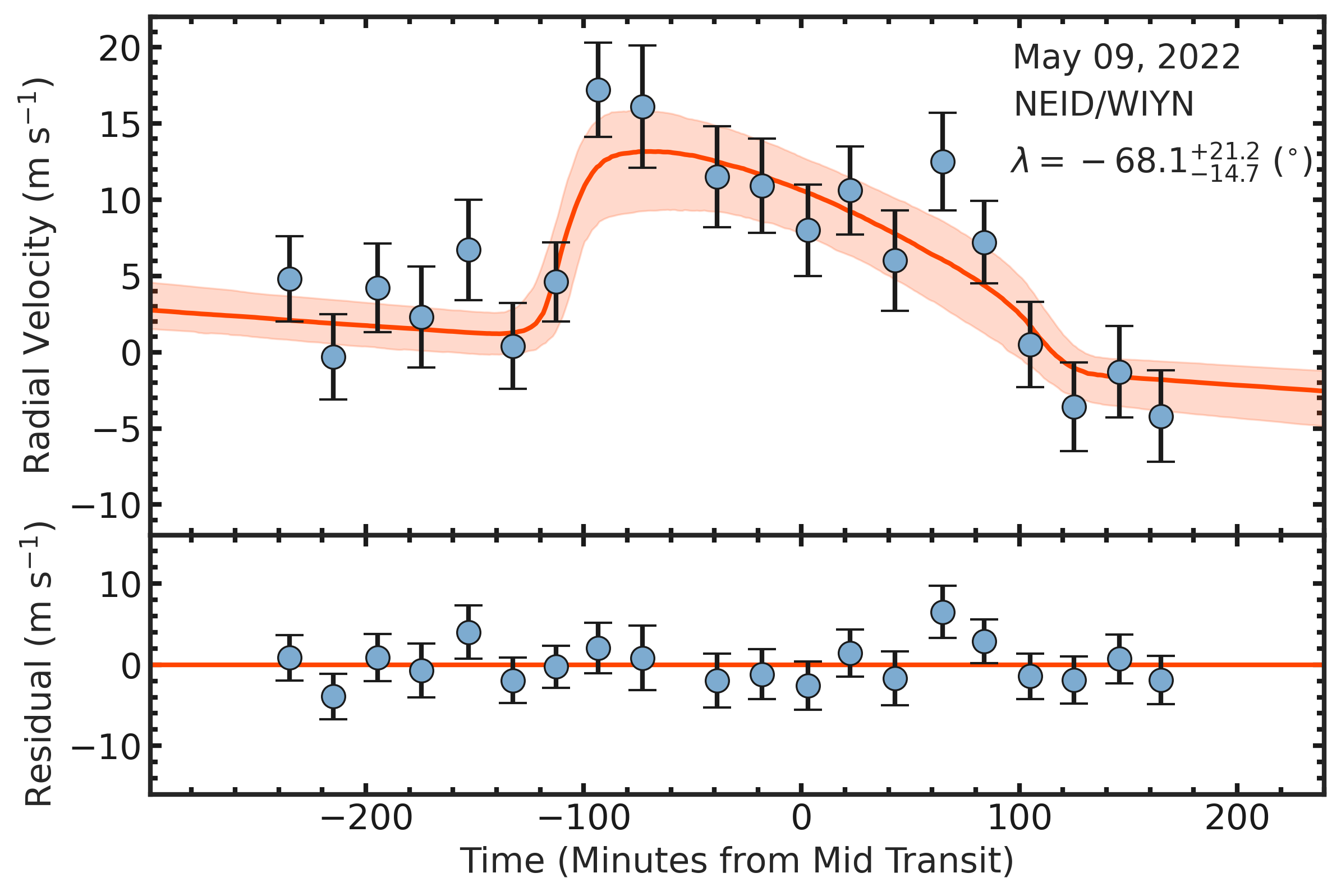}
    \caption{The measured Rossiter-McLaughlin effect for TOI-1842 b suggests a significant misaligned orbit with a sky-projected obliquity of $\lambda=-68.1_{-14.7}^{+21.2} \, \degree$. The blue points with black error bars represent the measured in-transit radial velocities and their errors. The median model of the R-M effect is shown in red, with the corresponding $1 \, \sigma$ uncertainty shown in lighter red. This model was obtained from a global fit of all available radial velocity and transit data.}
    \label{fig:rm_fit}
\end{figure}

We used \texttt{allesfitter} \citep{gunther2020allesfitter} to model the sky-projected spin-orbit angle, $\lambda$, for TOI-1842b by performing a simultaneous global fit on light curves from \emph{TESS} Sectors 23 and 50, in-transit RVs from NEID, and published out-of-transit RVs from MINERVA and NRES obtained from \citet{wittenmyer2022toi1842}. 

The fitted parameters in our analysis are listed in Table \ref{table:results}. Each parameter was initialized with uniform priors, while initial guesses for $P$, $T_0$, $\rp / \rstar$, $(\rstar + \rp )/ a$, $K$, $\ecosw$, and $\esinw$ were obtained from the TOI-1842b planet discovery paper \citep{wittenmyer2022toi1842}. Additionally, we fitted the transformed limb darkening coefficients for \emph{TESS} ($q_{\rm 1:TESS}$, $q_{\rm 2:TESS}$) and NEID ($q_{\rm 1:NEID}$, $q_{\rm 2:NEID}$), all of which were initialized with a value of $0.5$ and sampled from a uniform distribution between 0 and 1. For each spectrograph used, a free parameter was introduced to account for potential radial velocity offsets. To account for instrument-specific effects, jitter terms were modeled and added in quadrature to the uncertainties. The bounds on the free parameter $\lambda$ were established as $-180\degr$ to $+180\degr$.

We used the Affine Invariant Markov Chain Monte Carlo (MCMC, \citealt{Goodman2010}) method (implemented with \texttt{emcee} \citep{foremanmackey2013}) with 100 walkers, each with 400,000 accepted steps, to sample the posterior distributions of all model parameters. The best-fit parameters and their 1$\, \sigma$ uncertainties are reported in Table \ref{table:results} and are typically within $1 \, \sigma$ agreement with the values presented in \citet{wittenmyer2022toi1842}, except for $\teff$, where we derive a lower value. Figure \ref{fig:rm_fit} shows the best-fit R-M model from our global fit, along with the corresponding residuals. The analysis reveals that TOI-1842b is misaligned, with sky-projected spin-orbit angle $\lambda=-68.1_{-14.7}^{+21.2} \degree$.

To derive the stellar spin velocity and constrain the spin-orbit angle along the line of sight, we performed a periodogram analysis \citep{Zechmeister2009} on the TOI-1842 \emph{TESS} light curve (Sectors 23 and 50) with the transits masked out. This resulted in a rotational period (\prot) of $11.350 \pm  0.006$ days with a false-alarm probability (FAP) of less than 0.001. We adopted $\prot$ to be $11.350 \pm 1.135$ d, since the latitudinal differential rotation establishes a lower bound for the precision of the measurements for $\prot$ at 10\% \citep{Epstein2014,Aigrain2015}. Based on this value, the stellar equatorial rotation velocity is calculated as $v=\frac{2\pi\rstar}{\prot}=9.0\pm1.0 \,\rm km/s$. To derive the stellar inclination, we employ MCMC method on $\rstar$, $P_{\rm rot}$, and cos$i$ to account for interdependent variables $v$ and $\vsini$ \citep{Masuda2020stincl,Hjorth2021}. Gaussian priors on \rstar, \prot, and $v\sqrt{(1-cos^{2}i_{*})}$ were adopted. The likelihood function was taken to be

\begin{equation}
\begin{aligned}
\mathcal{L} &= \left(\frac{R_{\star} / R_{\odot} - 2.02}{0.07}\right)^{2} + \left(\frac{P_{\mathrm{rot}} - 11.350 \mathrm{~d}}{1.135 \mathrm{~d}}\right)^{2} \\
&\phantom{=} + \left(\frac{v\sqrt{(1-\cos^{2}i_{*})} - 6.03 \mathrm{~km} / \mathrm{s}}{0.89 \mathrm{~km} / \mathrm{s}}\right)^{2}.
\end{aligned}
\end{equation}

We ran the MCMC process for 20,000 steps and 100 walkers, obtaining 50 independent samples, which converged. The resulting stellar inclination is $46.4^{+12.3}_{-10.1}\degree$. Then, the true stellar obliquity $\psi$ can be derived through the equation below \citep[eq. 1]{albrecht2022obliq_review},

\begin{equation}
    \cos \psi=\cos i_{*} \cos i+\sin i_{*} \sin i \cos \lambda
\end{equation}
where $i_{*}$ is the stellar inclination and $i$ is the orbital inclination. The resulting 3D obliquity ($\psi$) of TOI 1842 was determined to be $\psi=73.3^{+16.3}_{-12.9}\degree$, indicating its misalignment.



\begin{deluxetable*}{lccc}
\tablecaption{Priors and posteriors for the TOI-1842 b system global fitting.  \label{table:results}}
\tabletypesize{\scriptsize}
\tablehead{
  \colhead{ } & \colhead{NEID Spectrum}  & \colhead{MIST+SED} \\
 \colhead{} & \colhead{iSpec}  & \colhead{EXOFASTv2} 
}
\tablewidth{300pt}
\startdata
\multicolumn{3}{l}{Stellar Parameters:}\\
~~~~$\mstar$ ($\mathrm{M}_\odot$) & - & $1.45^{+0.07}_{-0.14}$ & \\
~~~~$\rstar$ ($\mathrm{R}_\odot$) & - & $2.03\pm0.07$ &  \\
~~~~$\logg$ (cgs) & $4.19\pm0.28$ & $3.98^{+0.04}_{-0.05}$ &  \\
~~~~$[M/H]$ (dex) & $0.09\pm0.10$ & $0.27^{+0.13}_{-0.15}$ & \\
~~~~$\teff$ (K) & $5931\pm174$ & $6033^{+95}_{-93}$ &  \\
~~~~$\vsini$ (km/s) & $6.03\pm0.89$ &- & \\
\hline
\hline
&Priors for global fit&Global fit 1: NEID \\
\hline
\multicolumn{3}{l}{Stellar Parameters:}\\
~~~~$\vsini$ (km/s)  &  $\mathcal U(4.3; 0.0; 20)$    & $6.21_{-1.49}^{+3.64}$      \\  
~~~~ $\prot$ (days) & - & $11.350 \pm 1.135$ \\
~~~~ $i_{*}$ (deg) & - & $46.4^{+12.3}_{-10.1}$ \\
~~~~ $\psi$ (deg) & - & $73.3^{+16.3}_{-12.9}$ \\
\multicolumn{3}{l}{Planetary Parameters:}\\
\multicolumn{3}{l}{~~~~\textbf{TOI-1842b:}}\\
~~~~$\lambda_{\rm b}$ (deg) &  $\mathcal U(0; -180; +180)$           & $-68.1_{-14.7}^{+21.2}$     \\    
~~~~$P_{\rm b}$ (days) &  $\mathcal U(9.5739; 8.5739; 10.5739)$   & $9.5740\pm0.0001$   \\               
~~~~$R_{P;b} \ (R_J)$  & -   & $1.06_{-0.06}^{+0.07}$     \\      
~~~~$M_{P;b} \ (M_J)$  & -   & $0.19_{-0.04}^{+0.06}$     \\ 
~~~~$T_{0;b} \ (\rm BJD) - 2459300 $     &    $\mathcal U(25.871;24.871;26.871)$ 	& $25.871 \pm 0.005$ \\
~~~~$i_{b}$ (deg)   & -   & $87.0_{-2.2}^{+1.7}$    \\  
~~~~$e_{b}$ & -           & $0.13_{-0.09}^{+0.16}$     \\             
~~~~$\omega_{b}$ (deg)  & -    & $100.5_{-37.5}^{+93.2}$    \\       
~~~~$\cos{i_{\rm b}}$   &  $\mathcal U(0.0; 0.0; 1.0)$  & $0.05_{-0.03}^{+0.04}$         \\           
~~~~$K_{\rm b}$ ($\rm m \ s^{-1}$) &  $\mathcal U(15.9; 0.0; 1000.0)$    & $17.3\pm3.0$      \\            
~~~~$\rp / \rstar$ &  $\mathcal U(0.052; 0.0; 1.0)$              & $0.0540_{-0.0026}^{+0.0032}$  \\      
~~~~$(\rstar + R_{ \rm b} ) / a_{\rm b}$ &  $\mathcal U(0.028; 0.0; 1.0)$   & $0.08\pm0.02$  \\  
~~~~$\sqrt{e_b} \cos{\omega_{\rm b}}$ &   $\mathcal U(0.0; -1.0; 1.0)$  & $-0.01_{-0.17}^{+0.19}$    \\ 
~~~~$\sqrt{e_b} \sin{\omega_{\rm b}}$ &  $\mathcal U(0.0; -1.0; 1.0)$ & $0.3_{-0.3}^{+0.2}$    \\         
\multicolumn{3}{l}{Transformed limb darkening coefficients:}\\
~~~~$q_{\rm 1:NEID}$ & $\mathcal U(0.5;0;1)$ & $0.39_{-0.28}^{+0.37}$ \\
~~~~$q_{\rm 2:NEID}$ & $\mathcal U(0.5;0;1)$  & $0.39_{-0.28}^{+0.37}$ \\
~~~~$q_{\rm 1:\emph{TESS}}$ & $\mathcal U(0.5;0;1)$   & $0.47_{-0.22}^{+0.31}$  \\
~~~~$q_{\rm 2:\emph{TESS}}$ &  $\mathcal U(0.5;0;1)$  & $0.34_{-0.22}^{+0.33}$ \\
\multicolumn{3}{l}{Physical limb darkening coefficients:}\\
~~~~$u_{\rm 1:NEID}$     &  - &  $0.40_{-0.29}^{+0.49}$                       \\
~~~~$u_{\rm 2:NEID}$     &  - & $0.10_{-0.36}^{+0.40}$                      \\
~~~~$u_{\rm 1:\emph{TESS}}$  &  - & $0.46_{-0.30}^{+0.31}$    \\
~~~~$u_{\rm 2:\emph{TESS}}$ & - & $0.21_{-0.39}^{+0.37}  $      
\enddata 
\end{deluxetable*}

\section{Discussion}
\label{section:discussion} 

It has been observed that hot Jupiters orbiting low-mass ($\mstar<1.2\, \msun$), cool ($\teff \lesssim 6250 \, \rm K$) stars tend to be spin-orbit aligned, while those orbiting hot, massive stars show a wider range of spin-orbit misalignments \citep{winn2010hot, schlaufman2010evidence,albrecht2012obliquities, wang2021aligned}. 

This trend has been attributed to tides, with the explanation being that cool stars with deep convective envelopes and slower rotation rates, below the Kraft break ($\teff \lesssim 6250 \, \rm K$, \citealt{Kraft1967}), undergo faster tidal dissipation, resulting in their realignment with respect to their companion hot Jupiters' orbits \citep{albrecht2012obliquities}. In contrast, hot, massive star systems are thought to retain their primordial obliquity due to slower tidal dissipation \citep{winn2010hot,albrecht2012obliquities,winn2015,albrecht2022obliq_review}. 

Additionally, previous works have suggested that Saturn-mass ($M_p\sim0.2-0.4 \, \mj$) and, by extension, sub-Saturn planets, may be more commonly misaligned than higher-mass Jovian planets \citep{schlaufman2010evidence,sanchis2013kepler63,wang2018kepler9,anisman2020wasp117,dong2022toi-1268,rice2022WJaligned}. Sub-Saturns may be relatively susceptible to planet-planet scattering \citep{rasio1996dynamical, raymond2010planet} and/or secular misalignment mechanisms \citep[e.g.\ ][]{petrovich2020disk}. As sub-Saturns are less massive than Jovian planets, their host stars can less easily realign, as tidal dissipation timescales scale with $\mplanet^{-2}$ \citep{albrecht2012obliquities,albrecht2022obliq_review}. 

However, recent research has found that warm Jupiters tend to be preferentially aligned \citep{rice2022WJaligned}. Given that warm Jupiters have longer orbital periods and are ``tidally detached'', meaning they cannot tidally realign their host star on the same timescale as the lifetime of the system, this challenges tidal dissipation as the origin for the alignment of warm Jupiters and suggests that they may be primordially spin-orbit aligned \citep{rice2022WJaligned,albrecht2022obliq_review,davies2019disk_aligned}.

One possibility is that hot and warm Jupiters may have formed through distinct mechanisms, with warm Jupiters forming through a more tranquil process and being initially aligned with their host star's equator, while hot Jupiters form through a more tumultuous process and are therefore misaligned \citep{dawson2018origins,rice2022HJobliqDistr}. It is only cool, low-mass stars with orbiting hot Jupiters that may have undergone tidal realignment, leading to the observed trend \citep{winn2010hot,anderson2021possible,albrecht2022obliq_review}.

We, however, propose a potential alternative for the observed relation between $\mstar$/$\teff$ and their obliquities. Our framework is based on the ability of systems to produce multiple compact planets and \textbf{have} their obliquity excited through post-disk planet-planet interactions. The currently observed stellar obliquity distribution may be a result of the planet-planet interactions, with tides playing a lesser role in altering obliquities.

Throughout this section, we define warm Jupiters as planets with $M_p > 0.3 \mj$ and star-planet separation $a/R_*>11$. Accordingly, we define hot Jupiters as their closer-in analogues with $M_p > 0.3 \mj$ and star-planet separation $a/R_*\leq11$, and we define Saturn-mass planets as those with $M_p<0.3 \, \mj$.

\begin{figure*}
    \centering
    \includegraphics[width=1.0\linewidth]{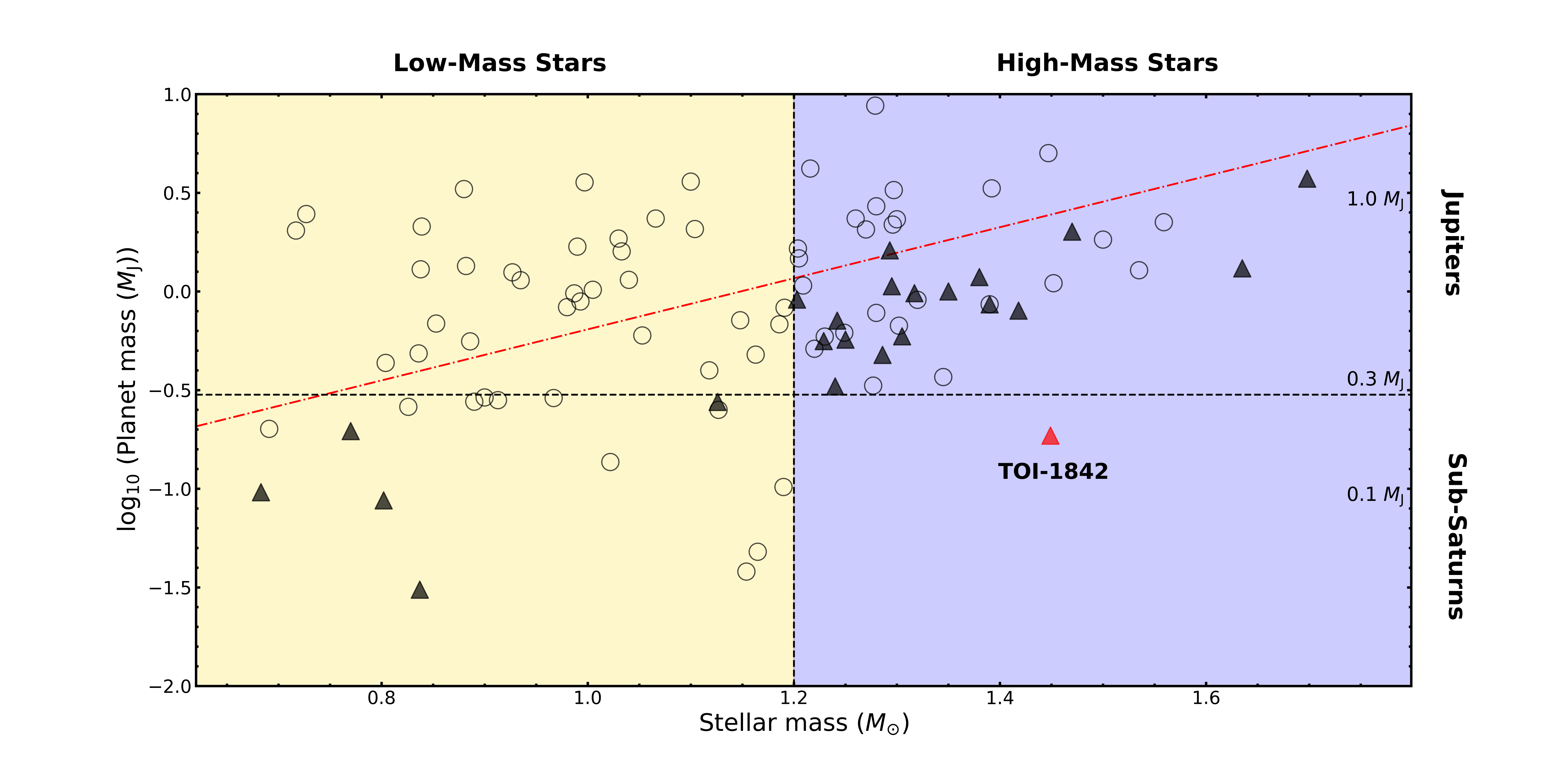}
    \caption{Stellar mass vs. planet mass for systems obtained from \citet{albrecht2022obliq_review} and the TEPCat catalogue \citep{southworth2011homogeneous}. Unfilled circles denote aligned planets, while triangles denote misaligned planets. TOI-1842 is plotted in red. The trend toward alignment that is confined to Jupiter-mass planets orbiting cool, low-mass stars with all other mass combinations showing mixed alignment. Misaligned systems all fall below the red dashed line representing the $1\sigma$ upper error to a linear fit performed on the misaligned systems.}
    \label{fig:mass_mass}
\end{figure*}

It has been demonstrated that disk and stellar mass are correlated, with disk-to-stellar mass ratios on the order of a few percent \citep{williams2011disks,andrews2013disks,andrews2020disks}. Consequently, massive stars are associated with more massive protoplanetary disks \citep{andrews2013mass,pascucci2016steeper} capable of forming multiple Jupiter-mass planets \citep{johsnon2010giant_occurence,ghezzi2018giant_occurence,yang2020architecture}. Cool, low-mass stars have lower-mass disks that may only form a single Jupiter-mass planet \citep{andrews2013mass,ansdell2016disks,pascucci2016steeper,dawson2018origins, yang2020architecture}.
Thus, the observed $\mstar$/$\teff$ and obliquity relationship could be explained as \textbf{an indication that} hot stars are more capable of producing multiple Jupiters that interact and result in a wide range of obliquities. On the other hand, cool, low-mass stars lack the material to form multiple Jupiter-mass planets; thus these systems stay in a stable, aligned configuration with a single Jupiter. This can be seen in the top panels of Figure \ref{fig:mass_mass}.


Our hypothesis applies to low-mass planets around both cool and hot stars, with a wide range of spin-orbit angles expected for both types of stars, since both are capable of producing multiple low-mass planets, such as sub-Saturns. The bottom panels of Figure \ref{fig:mass_mass} show that sub-Saturns around cool stars span a wide range of stellar obliquities, in support of our hypothesis. However, TOI-1842b is the first sub-Saturn orbiting a hot, massive star with an R-M measurement. Its orbital misalignment supports our explanation. Further measurements in this population are needed to make a strong conclusion. 

To further explain our hypothesis, we present the four quadrants of star and planet mass and their observed obliquities in Figure \ref{fig:mass_mass}. Throughout this section and in Figure \ref{fig:mass_mass}, we define ``misaligned'' systems as those with $|\lambda|>10^{\circ}$ and $\lambda$ differing from $0\degree$ at a $3 \, \sigma$ level, following the definition in \citet{wang2022wasp148}.

\begin{itemize}
    \item \textbf{Jupiters around Cool Stars:} Cool, low-mass stars tend to have less massive protoplanetary disks, which may only be capable of forming a single Jupiter-mass planet \citep{andrews2013mass,ansdell2016disks,pascucci2016steeper,dawson2018origins, yang2020architecture}. In these cases, there is no potential perturber in the same systems with sufficient mass to excite the spin-orbit angle of the lone Jupiter, resulting in aligned Jupiters around cool, low-mass stars as seen in the top left panel of Figure \ref{fig:mass_mass}.

    \item \textbf{Jupiters around Hot Stars:} Hot, massive stars tend to have more massive protoplanetary disks that are capable of forming multiple Jupiter-mass planets \citep{johsnon2010giant_occurence,ghezzi2018giant_occurence,yang2020architecture}. The presence of multiple Jupiters presents the opportunity for interactions, such as Jupiter-Jupiter scattering \citep{rasio1996dynamical, chatterjee2008scattering} or secular interactions \citep{wu2011secularchaos, petrovich2015migrationKL,naoz2011secular}, that can excite the spin-orbit angles of the Jupiters. As a result, Jupiters around hot, massive stars may be found to be spin-orbit misaligned due to the planet-planet interactions with other massive planets initially present in the same systems. This framework for Jupiters around hot, massive stars can be seen in the top right panel of Figure \ref{fig:mass_mass} and matches the $\teff$ vs. obliquity relationship.
    
    \item \textbf{Sub-Saturns around Cool Stars:} Despite the fact that cool, low-mass stars tend to have less massive protoplanetary disks, it is still possible for these disks to initially form multiple lower-mass planets, such as sub-Saturns. This scenario is more similar to the presence of multiple Jupiters around hot stars, as both situations involve the potential for interactions among compact multiple planets initially present in the same systems through mechanisms like planet-planet scattering or secular interactions, which can excite their spin-orbit angles. This can be seen through the mixed obliquity distribution for sub-Saturns around cool stars in the bottom left panel of Figure \ref{fig:mass_mass}.
    
    \item \textbf{Sub-Saturns around Hot Stars:} Given the similarity of the presence of multiple lower-mass planets around cool, low-mass stars and multiple Jupiters around hot, massive stars in terms of the potential for planet-planet interactions and exciting their spin-orbit angles, it is reasonable to expect that a significant fraction of low-mass planets around hot, massive stars may also exhibit spin-orbit misalignment. Once again, this is due to post-disk planet-planet interactions in a proto-planetary disk capable of producing multiple low-mass planets.

    However, current observations in this area are limited, with TOI-1842b being the first measurement in this regime. Nevertheless, the misalignment of TOI-1842b supports this hypothesis, as shown in the bottom right panel of Figure \ref{fig:mass_mass}. Further observations of low-mass planets around hot, massive stars will provide additional evidence to examine this region of parameter space. Doing so will be vital for further tests of this hypothesis.
    
\end{itemize}

There appears to be an upper limit when looking at the misaligned systems in planet mass vs.\ stellar mass space. We place an empirical upper limit on these systems by performing a linear fit to these misaligned systems. The $1\sigma$ upper error line of this fit is plotted as a dashed red line Figure\ref{fig:mass_mass}. We note that all misaligned systems in the data fall below this $1\sigma$ line. Above this line, we expect systems to be aligned as the planet-to-stellar mass ratio is higher and it is unlikely these systems can produce multiple planets capable of interacting to excite obliquity. Below this line, we expect mixed alignment as the ratio of planet-to-stellar mass is low enough to produce multiple planets for which obliquity excitation works in the presented hypothesis.

Although our framework proposes an alternative explanation for the observed spin-orbit angle distributions, the sharp cutoff of the obliquity distribution at the Kraft break (see top panel of Figure~\ref{fig:mass_mass} in this paper, and Figure 8 in \citealt{albrecht2022obliq_review}) suggests that tidal dissipation could play a role in shaping the obliquity ranges of cool stars hosting hot Jupiters. Our hypothesis complements tidal theory and implies that initial degrees of obliquity for these cool stars may be smaller than those for hot stars. If our framework is correct, we anticipate that the obliquity distribution of warm Jupiters will show an increasing spread with a rising stellar mass. This is because the original obliquity distribution would show larger scatter with stellar mass due to an increased probability of multi-planet interactions that can  excite obliquity. For warm Jupiters around hot stars, this obliquity should not be tidally damped, since the timescale for tidal re-alignment for warm Jupiters is expected to be longer than the lifetime of the system.

\section{Acknowledgements}
\label{section:acknowledgements}
We thank Lauren Weiss and Ji Wang for their insightful discussion at GLEAM 2022. Additionally, we thank Francisco Aros for his discussion during an Indiana University Astronomy Department Tea Talk. We also acknowledge the contributions of Brandon Radzom and Armaan Goyal for conversations regarding statistical methods. We are also grateful for general discussion with Jiayin Dong related to this work. We also thank Heidi Schweiker, Sarah E. Logsdon, and the NEID Queue Observers and WIYN Observing Associates for their skillful execution of our NEID observations, as well as the The NEID-SpecSoft Team for their data reduction pipeline. M.R. and S.W. thank the Heising-Simons Foundation for their generous support. This research was supported in part by Lilly Endowment, Inc., through its support for the Indiana University Pervasive Technology Institute.

\software{\texttt{numpy} \citep{oliphant2006guide, walt2011numpy, harris2020array}, \texttt{matplotlib} \citep{hunter2007matplotlib}, \texttt{pandas} \citep{mckinney2010data}, \texttt{scipy} \citep{virtanen2020scipy}, \texttt{allesfitter} \citep{gunther2020allesfitter}, \texttt{emcee} \citep{foremanmackey2013}}

\facility{WIYN/NEID}

\bibliography{bibliography}
\bibliographystyle{aasjournal}

\end{document}